\documentclass[preprint,showpacs,showkeys,amsmath,amssymb]{revtex4}
\usepackage{graphicx} 
\usepackage{color}
\usepackage{ulem} 
\newcommand{\prcaco}{Pr$_{1-x}$Ca$_{x}$CoO$_3$}

\newcommand{\lasrco}{La$_{1-x}$Sr$_{x}$CoO$_3$}

\newcommand{\ndsrco}{Nd$_{1-x}$Sr$_{x}$CoO$_3$}

\newcommand{\lasrtri}{La$_{0.7}$Sr$_{0.3}$CoO$_3$}

\newcommand{\prcatri}{Pr$_{0.7}$Ca$_{0.3}$CoO$_3$}
\newcommand{\ndsrtri}{Nd$_{0.7}$Sr$_{0.3}$CoO$_3$}
\newcommand{\ndcatri}{Nd$_{0.7}$Ca$_{0.3}$CoO$_3$}
\newcommand{\prycatri}{(Pr$_{1-y}$Y$_{y}$)$_{0.7}$Ca$_{0.3}$CoO$_3$}
\newcommand{\laco}{LaCoO$_3$}
\newcommand{\srco}{SrCoO$_3$}

\newcommand{\st}{$\rm ^o$}
\newcommand{\stc}{$\rm ^oC$}
\newcommand{\etal}{\textit{et~al.}}
\newcommand{\jkmol}{J$\cdot$mol$^{-1}$K$^{-1}$}

\begin{document}
\sloppy
\title{
 Ground state properties of the mixed-valence cobaltites
 Nd$_{0.7}$Sr$_{0.3}$CoO$_3$, Nd$_{0.7}$Ca$_{0.3}$CoO$_3$ and Pr$_{0.7}$Ca$_{0.3}$CoO$_3$}
\author{Z. Jir\'{a}k}
\author{J. Hejtm\'{a}nek}
\author{K. Kn\'{\i}\v{z}ek}
\author{M. Mary\v{s}ko}
\author{P. Nov\'{a}k}
\author{E. \v{S}antav\'{a}}
\affiliation{Institute of Physics ASCR, Cukrovarnick\'a 10, 162 00 Prague 6, Czech Republic.}
\author{T. Naito}
\author{H. Fujishiro}
\affiliation{Faculty of Engineering, Iwate University, 4-3-5 Ueda, Morioka 020-8551, Japan.}
%
\begin{abstract}
The electric, magnetic, and thermal properties of three perovskite cobaltites with the same 30\%
hole doping and ferromagnetic ground state were investigated down to very low temperatures. With
decreasing size of large cations, the ferromagnetic Curie temperature and spontaneous moments of
cobalt are gradually suppressed - $T_C=130$~K, 55~K and 25~K and $m~=~0.68~\mu_B$, 0.34~$\mu_B$
and 0.23~$\mu_B$ for Nd$_{0.7}$Sr$_{0.3}$CoO$_3$, Pr$_{0.7}$Ca$_{0.3}$CoO$_3$ and
Nd$_{0.7}$Ca$_{0.3}$CoO$_3$, respectively. The moment reduction with respect to moment of the
conventional ferromagnet La$_{0.7}$Sr$_{0.3}$CoO$_3$ ($T_C=230$~K, $m~=~1.71~\mu_B$) in so-called
IS/LS state for Co$^{3+}$/Co$^{4+}$, was originally interpreted using phase-separation scenario.
Based on the present results, mainly the analysis of Schottky peak originating in Zeeman splitting
of the ground state Kramers doublet of Nd$^{3+}$, we find, however, that ferromagnetic phase in
Nd$_{0.7}$Ca$_{0.3}$CoO$_3$ and likely also Pr$_{0.7}$Ca$_{0.3}$CoO$_3$ is uniformly distributed
over all sample volume, despite the severe drop of moments. The ground state of these compounds is
identified with the LS/LS-related phase derived theoretically by Sboychakov~\textit{et~al.} [Phys.
Rev. B \textbf{80},~024423~(2009)]. The ground state of Nd$_{0.7}$Sr$_{0.3}$CoO$_3$ with an
intermediate cobalt moment is inhomogeneous due to competing of LS/LS and IS/LS phases. In the
theoretical part of the study, the crystal field split levels for $4f^3$ (Nd$^{3+}$), $4f^2$
(Pr$^{3+}$) and $4f^1$ (Ce$^{3+}$ or Pr$^{4+}$) are calculated and their magnetic characteristics
are presented.
\end{abstract}
\maketitle
\pagestyle{myheadings} \markright{(PrNd)$_{0.7}$(SrCa)$_{0.3}$CoO$_3$. Date: \today}
\section{Introduction}

In the perovskite cobaltites, two prototypical behaviors can be distinguished. The first one is
associated with spin transition, or spin-state crossover of Co$^{3+}$ ions in \laco\ and its rare
earths analogs, while the second one is manifested by robust ferromagnetic metallic ground state
that is observed in the mixed-valence \lasrco\ systems above a critical doping value of $x=0.22$
and exists up to the formally pure Co$^{4+}$ end compound \srco\
\cite{RefCaciuffo1999PRB59_1068,RefWu2003PRB67_174408,RefLong2011JPCM23_245601}.

It is well established that the ground state of \laco\ is based on non-magnetic LS (low spin,
$t_{2g}^6$) states. With temperature increasing above $\sim 40$~K the energetically close HS (high
spin, $t_{2g}^4 e_g^2$) species start to be populated by thermal excitation. The process is
readily seen in the course of magnetic susceptibility or in anomalous terms of lattice expansion
due to ionic size of HS notably larger than LS one (see \textit{e.g.}
\cite{RefKnizek2006JPCM18_3285}). These experiments show that HS population increases gradually
with steepest rate at $\sim 80$~K, and is practically saturated above 150~K, making about
$40-50$\%. Strong HS/LS nearest neighbor correlations or even short-range ordering are anticipated
in that phase \cite{RefGoodenough1958JPCS6_287,RefBari1972PRB5_4466,RefKnizek2009PRB79_014430GGA}.
At still higher temperature the ordering melts, which is accompanied by a drop of the electrical
resistivity and thermopower values at about 530~K, reminding the I-M transition. The character of
the high-temperature phase is not yet recognized - IS Co$^{3+}$ states (intermediate spin,
$t_{2g}^{5} e_g^1$) have been tentatively suggested
\cite{RefKyomen2005PRB71_024418,RefJirak2008PRB78_014432,RefKnizek2009PRB79_014430GGA}, but very
recent LDA+DMFT calculations did not reveal any significant amount of IS Co$^{3+}$ in the
fluctuating mixture of spin states in \laco\ at high temperatures
\cite{RefKrapek2012PRB86_195104}.

The metallic conductivity in \lasrco\ systems with stable ferromagnetic ground state is related to
fast valence fluctuations that involve various $3d^N$ ($N$=5,~6,~7) states on cobalt sites
\cite{RefKunes2012PRL109_117206}. In particular for \lasrtri, the dominant population are the
IS/LS states for Co$^{3+}$/Co$^{4+}$, so that the electronic structure can be approximated as
$t_{2g}^{5} \sigma^{*0.7}$, where $\sigma^*$ is an antibonding band of the $e_{g}$ parentage. This
yields spontaneous moment of 1.70~$\mu_B$ in agreement with 1.71(2)~$\mu_B$ determined by neutron
diffraction (Curie temperature $T_C=230$~K)
\cite{RefCaciuffo1999PRB59_1068,RefWu2003PRB67_174408}. The spontaneous moment increases with
further hole doping, but most of the published data for $x>0.5$ are influenced by severe oxygen
deficiency of the samples. Nevertheless, nearly pure Co$^{4+}$ system \srco\ has been prepared
under very high pressures, yielding the moment of 2.5~$\mu_B$ ($T_C=305$~K)
\cite{RefLong2011JPCM23_245601}. This means that electrons are not depopulated from the
$e_{g}$-derived $\sigma^*$ band but from the more localized $t_{2g}$ levels, so that, in gross
simplification, the final electronic structure of \srco\ can be presented as $t_{2g}^{4}
\sigma^{*1}$ (see Ref.~\cite{RefKunes2012PRL109_117206} for a more detailed description).

As the transition range between \laco\ and \lasrtri\ is concerned, significant data have been
obtained in the single crystal studies of He~\etal\
\cite{RefHe2009EPL87_027006,RefHe2009PRB80_214411}. The \lasrco\ systems above $x=0.22$ show
characteristics of conventional ferromagnets. The ferromagnetic-paramagnetic (FM-PM) transition is
manifested by large $\lambda$ peak in the specific heat and by the presence of sharp critical
scattering peak in small-angle neutron scattering at $T_C$. As shown in recent re-investigation of
critical exponents $\beta$, $\gamma$ and $\delta$ by Khan~\etal, the transition is unambiguously
of second order, characterized by scaling behavior that belongs to universality class of 3D
Heisenberg model \cite{RefKhan2010PRB82_064422}. On the contrary, in single crystal samples below
$x=0.22$ the anomalies at $T_C$ are absent or smeared out, pointing to a much more complex
temperature behavior including the magnetic/electronic phase separation. The magnetic
inhomogeneity of low-doped \lasrco\ has been proved as well in the inelastic neutron diffraction
by Phelan~\etal\ \cite{RefPhelan2006PRL97_235501} or in NMR study by Smith~\etal\
\cite{RefSmith2008PRB78_092201}.

The two regimes of behavior in \lasrco, separated by the critical doping $x=0.22$, have been
interpreted theoretically by Sboychakov~\etal\ \cite{RefSboychakov2009PRB80_024423}. Based on a
fermionic model of Hubbard type, two possible FM ground states that may coexist on a nanoscopic
scale are found.  One is derived from the phase characterized by LS/LS state for
Co$^{3+}$/Co$^{4+}$, in which only few Co$^{3+}$ $t_{2g}$ electrons promote to itinerant $e_{g}$
levels with increasing strontium doping. The second phase is IS/LS state for Co$^{3+}$/Co$^{4+}$,
\textit{i.e.} the already mentioned $t_{2g}^{5} \sigma^*$ phase, which becomes dominant above the
critical composition $x_c \sim 0.20$. The model phase diagram presented in
Ref.~\cite{RefSboychakov2009PRB80_024423} thus reproduce remarkably well the features actually
observed in the \lasrco\ system.

The properties of mixed-valence cobaltites can be further modified by a control of ionic size on
perovskite A sites. Lot of studies have been done on the \prcaco\ system that exists in a limited
range up to $x\sim0.55$ \cite{RefTsubouchi2002PRB66_052418,RefTsubouchi2004PRB69_144406}. The
substitution of smaller Pr$^{3+}$ and Ca$^{2+}$ ions for La$^{3+}$ and Sr$^{2+}$ ions causes
larger size misfit between the A and B sites of perovskite structure (Goldschmidt's tolerance
factor is reduced), which results in larger deviation of the Co-O-Co bond angles from ideal
180\st\ and, subsequently, in narrowing of the $e_{g}$-derived band. The reduced bandwidth is
detrimental not only for macroscopic conductivity but also for ferromagnetic interactions mediated
by itinerant $e_{g}$ electrons (the double-exchange). Ferromagnetism is thus suppressed and, at
the same time, the LS Co$^{3+}$ states are promoted due to larger $t_{2g}$-$e_{g}$ gap. As an
ultimate effect we may mention the peculiar behavior of \prcaco\ systems within a range
$x=0.50-0.55$, which exhibit a sharp first-order transition to the low-temperature highly
resistive and weakly magnetic state. It is of interest that the transition is associated with a
charge transfer between the cobalt and praseodymium sites, resulting in a stabilization of
formally Pr$^{4+}$ states - see \textit{e.g.}
\cite{RefHerreroMartin2012PRB86_125106,RefFujishiro2012JPSJ81_064709} and references therein. In
the region of 30\% electron hole doping, to which our study is focused, the pronounced effects of
competing ground states were reported for \prcatri, as well as for \ndcatri, and the
phase-separated nature of these compounds was guessed also from the magnetization values, which
are severely reduced compared to \lasrtri\ ones
\cite{RefKundu2004JMMM281_261,RefKundu2006JSSC179_923,RefKalinov2010PRB81_134427,RefElKhatib2010PRB82_100411}.

More direct information on the temperature dependence of magnetic/electronic phase separation was
recently provided by the neutron diffraction, small-angle scattering, and magnetometry study of
\prcatri\ by El-Khatib~\etal\ \cite{RefElKhatib2010PRB82_100411}. The study showed that FM
clusters (presumably hole-rich objects) were formed at a well-defined temperature $T^*\sim 250$~K,
while the hole-poor PM matrix was transformed to long-range FM order at much lower temperature
$T_C=70$~K. It is of interest that preformed clusters did not dissolve immediately, but were
preserved well below 70~K. This unusual coexistence of short- and long-range ordered FM phases
possessing different coercivities was further supported by observation of exchange-spring
behavior, which is known for artificial hard-soft magnetic composites
\cite{RefElKhatib2010PRB82_100411}. It should be noted that such a behavior refers to a
temperature range 50~-~70~K only. The aim of present study is to resolve what is the final
magnetic composition in \prcatri\ and related compounds at the lowest temperatures.

Our paper reports on the electric, magnetic, and heat capacity investigation of selected 30\%
doped systems, in decreasing order of tolerance factors \ndsrtri, \prcatri\ and \ndcatri. The
respective spontaneous moments are determined to 0.68~$\mu_B$, 0.34~$\mu_B$ and 0.23~$\mu_B$
($T_C=130$~K, 55 and 25~K). Such drop of magnetization compared to 1.71~$\mu_B$ in \lasrtri\
possessing robust IS/LS phase was previously interpreted within the phase separation scenario,
supposing that FM ordered regions of higher hole doping are imbedded in the non-magnetic matrix of
low hole doping
\cite{RefKundu2004JMMM281_261,RefKundu2006JSSC179_923,RefKalinov2010PRB81_134427,RefElKhatib2010PRB82_100411}.
The present results contradict such conjecture. Based on detailed characterization and novel
application of Kramers ions Nd$^{3+}$ as local probe of the magnetic ground state, we show that
the internal field formed by cobalt spins is uniformly distributed over the sample volume. The
presence of gross magnetic/electronic phase separation in \ndcatri\ or \prcatri\ at the lowest
temperatures is thus questioned. We argue that the ground state of these samples is in fact the FM
saturated LS/LS-derived phase suggested by Sboychakov~\etal\ \cite{RefSboychakov2009PRB80_024423}.

\section{Experimental}

Polycrystalline samples of \ndsrtri, \prcatri\ and \ndcatri\ were prepared using a solid-state
reaction. Raw powders of Pr$_6$O$_{11}$, Nd$_2$O$_3$, Y$_2$O$_3$, Co$_3$O$_4$, CaCO$_3$ and
SrCO$_3$ were weighted with proper molar ratios and ground using an agate mortar and pestle for
1~h. Mixed powders were calcinated at 1000\st C for 24~h in air. They were pulverized and ground.
Then they were pressed into pellets of 20~mm diameter and 4~mm thickness. Pellets were sintered at
1200\stc\ for 24~h in 0.1~MPa flowing oxygen gas. The measured densities of each sample were
greater than 90\% of the ideal density. Powder X-ray diffraction patterns were taken for each
sample using CuK$_\alpha$ radiation; the samples were confirmed to have a single phase
orthoperovskite ($Pbnm$) structure. The lattice parameters and volume \textit{per} f.u. actually
obtained are $a=5.364~$\AA, $b=5.409~$\AA, $c=7.599~$\AA, $V/Z=55.11~$\AA$^3$ for \ndsrtri,
$a=5.363~$\AA, $b=5.351~$\AA, $c=7.570~$\AA, $V/Z=54.32~$\AA$^3$ for \prcatri\ and $a=5.344~$\AA,
$b=5.342~$\AA, $c=7.549~$\AA, $V/Z=53.87~$\AA$^3$ for \ndcatri. The present values are in
agreement with the literature data for the same compounds
\cite{RefKundu2004JMMM281_261,RefMasuda2003JPSJ72_873}. As the important question of the oxygen
stoichiometry is concerned, practically ideal oxygen content $2.99\pm0.01$ is evidenced for the
\prcatri\ specimen by Rietveld refinement of the high resolution neutron diffraction data
\cite{RefNoteORNL}. The same seems to be valid for other two compounds based on the indirect
arguments, which follow from thermopower data presented below.

Electrical resistivity and thermoelectric power were measured using a four-probe method with a
parallelepiped sample cut from the sintered pellet. The electrical current density varied in
dependence on the sample resistivity between $10^{-1}$~A/cm$^2$ (metallic state) and
$10^{-7}$~A/cm$^2$ (insulating state). The measurements were made during cooling and warming the
sample. In the low temperature range, the close-cycle cryostat working down to $2-3$~K was used.
For the high-temperature experiments up to 1000~K the sample was placed on a ceramic sample holder
centered in the small tubular furnace with precisely controlled temperature. The standard
chromel-alumel thermocouples were used for monitoring of the temperature gradient around 5~K,
imposed across the sample by means of an additional small furnace.

The magnetic measurements were carried out using a SQUID magnetometer MPMS-XL  (Quantum Design) in
the temperature range $2-400$~K. The zero-field cooled (ZFC) and field cooled (FC)
susceptibilities were measured under different applied fields. The initial AC susceptibility
($H$~=~0) was studied in the frequency region 0.12~-~87.4~Hz using a driving AC field 3.9~Oe. The
virgin magnetization curves and hysteresis loops (-70~kOe, 70~kOe) in the ZFC regimes were
recorded for selected temperatures starting from $T=2$~K. In addition to it, the FC hysteresis
loops were measured after cooling the sample from $T=300$~K to $T=2$~K under an applied field
70~kOe.

The specific heat was measured by PPMS device (Quantum Design) using the two$-\tau$ model. The
data were collected on sample cooling. The experiments at very low temperatures (down to 0.4~K)
were done using the $^3$He option.

\section{Results}

\subsection{Electric transport}

\begin{figure}
\includegraphics[width=0.90\columnwidth,viewport=0 0 756 576,clip]{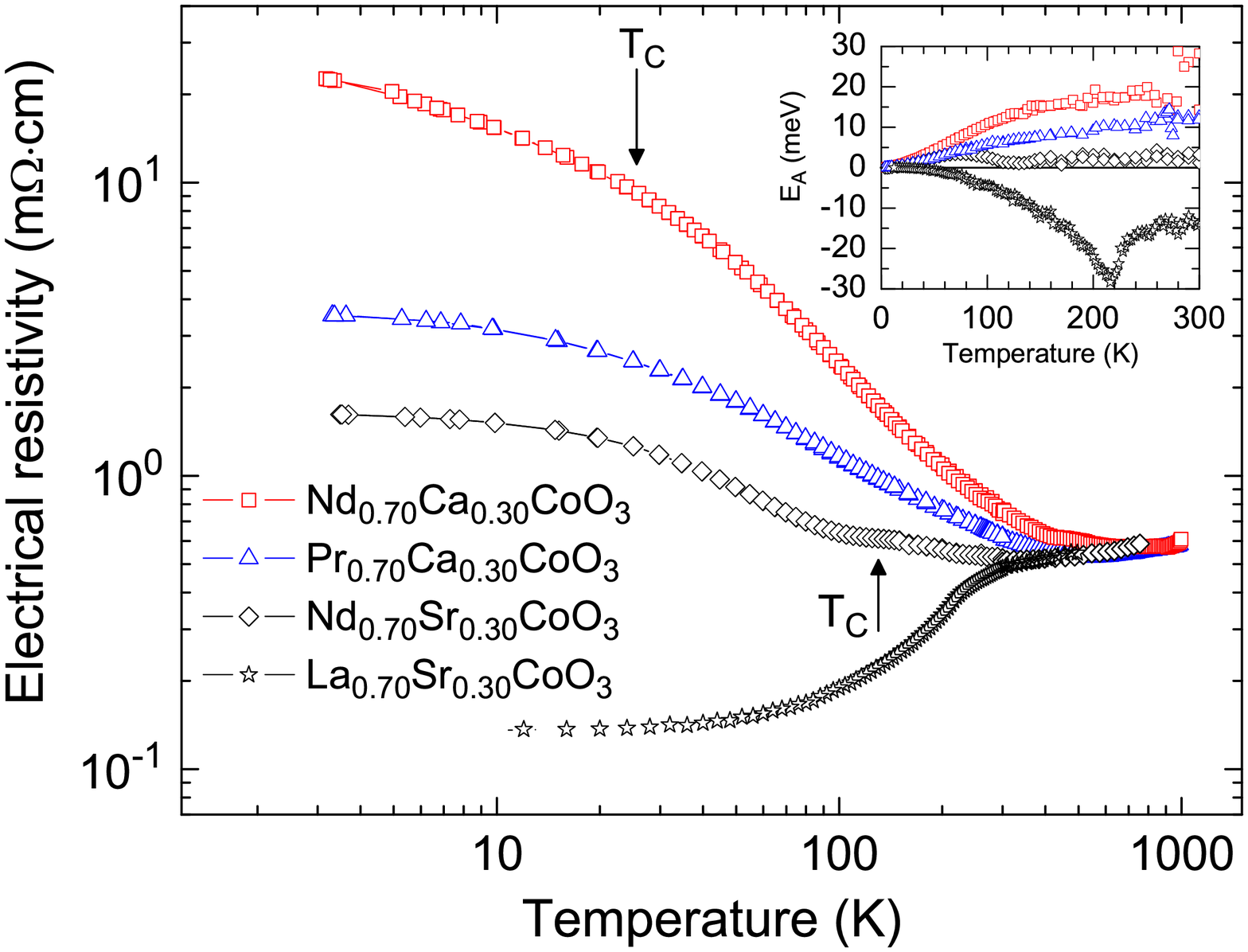}
\caption{(Color online) Resistivity of the \ndsrtri, \prcatri\ and \ndcatri\ ceramic samples.
Results on \lasrtri\ ceramics are added for comparison. The inset shows that, except for \ndcatri,
the apparent activation energy, defined as $E_A=k\cdot d(ln\rho)/d(1/T)$, is below the thermal
energy in the whole temperature range (the full line refers to $kT$).}
 \label{fig1}
\end{figure}

\begin{figure}
\includegraphics[width=0.90\columnwidth,viewport=0 0 756 576,clip]{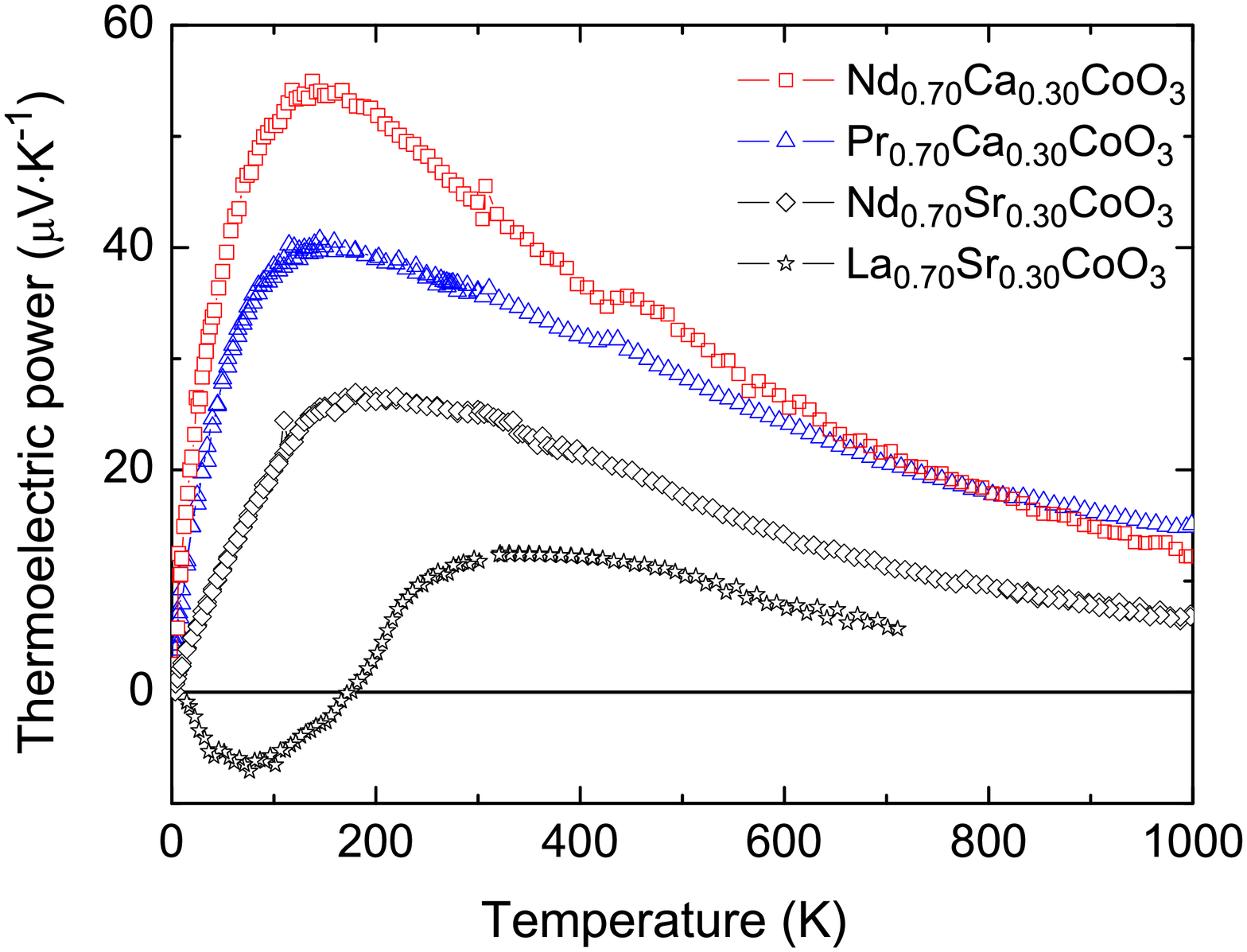}
\caption{(Color online) The thermopower data of \ndsrtri, \prcatri\ , \ndcatri\ and \lasrtri.}
 \label{fig2}
\end{figure}

Resistivity data obtained on ceramic samples \ndsrtri, \prcatri\ and \ndcatri\ are presented in
the $log-log$ plot in Fig.~\ref{fig1}. As the important question on metallic or insulating
character of these system is concerned, we argue that bulk properties are generally influenced by
the sample granularity, which seems to be also the present case. Namely, the measured resistivity
steadily increases with decreasing temperature but, instead of divergence at the lowest
temperatures, it extrapolates to finite values of about $1-10$~m$\Omega\cdot$cm at zero K, satisfying a
criterion for metallic ground state, $d(ln \rho)/d(ln T) \rightarrow 0$
\cite{RefMobius2000PHYSB284_1669}. Another signature for intrinsic metallicity is the apparent
activation energy, defined as $E_A=k\cdot d(ln\rho)/d(1/T)$. It does not exceed the thermal energy
$k_BT$, except for slightly enlarged $E_A$ values for \ndcatri\ above $T_C$ (see the inset of
Fig.~\ref{fig1}). On the other hand, the thermopower data are rather insensitive to presence of
grain boundaries. Seebeck coefficient is positive over the whole temperature range pointing to a
hole character of carriers (Fig.~\ref{fig2}). The low-temperature dependence is of linear
(metallic-like) type, but tends soon to a maximum at about 150~K, which achieves $\sim25, 40$ and
55~$\mu$V/K for \ndsrtri, \prcatri\ and \ndcatri, respectively. Then the thermopower slowly
decreases toward a plateau of 20~$\mu$V/K at high temperatures. Let us note that the maximum
values of thermopower are indirect but very sensitive indicators of the hole doping level and,
based on close agreement with data on samples oxygenated under 60~atm pressure in work of
Masuda~\etal\ \cite{RefMasuda2003JPSJ72_873}, they attest oxygen stoichiometry close to the ideal
one for our compounds.

The electric properties of \ndsrtri, \prcatri\ and \ndcatri\ differ substantially from the
behavior of analogously prepared \lasrtri\ ceramics, for which the data have been added to
Figs.~\ref{fig1} and \ref{fig2}. First, the resistivity in the \lasrtri\ sample shows a typical
metallic dependence with a resistivity drop below $T_C\sim220$~K. This becomes still more evident
in the graph of the apparent activation energy. More important signature of different ground state
is, however, the change of thermopower to negative values at low temperatures, in the case of
\lasrtri\, pointing to a dominant role of the electron carriers of $e_{g}$ parentage. On the other
hand, the above mentioned hole character of thermopower in \ndsrtri, \prcatri\ and \ndcatri\
suggests that the transport of $e_{g}$ carriers is impeded due to the band narrowing and/or
depopulation, and the $t_{2g}$ band becomes a prevalent conducting channel.

\subsection{Magnetic properties}

The magnetism in presently studied systems is governed by the cobalt spins and their exchange
interactions. The contribution of rare earths is manifested by Curie-like susceptibility with
effective moments that agree very well with free-ion values at intermediate and high temperatures,
$\mu_{eff}\sim 3.5\mu_B$ both for Nd$^{3+}$ and Pr$^{3+}$. However, a significant deviation from
this behavior is observed below $\sim 50$~K, where effects of crystal field splitting on 4$f$
shell become important. The low-temperature properties then depend critically on the character of the rare-earth ground state, which is Kramers doublet for Nd$^{3+}$ in the low-symmetry
crystal field of perovskite A sites, while the non-Kramers ions
Pr$^{3+}$ possess a singlet state. This issue is discussed in detail below in
Part C and Appendix.

\begin{figure}
\includegraphics[width=0.90\columnwidth,viewport=0 0 756 576,clip]{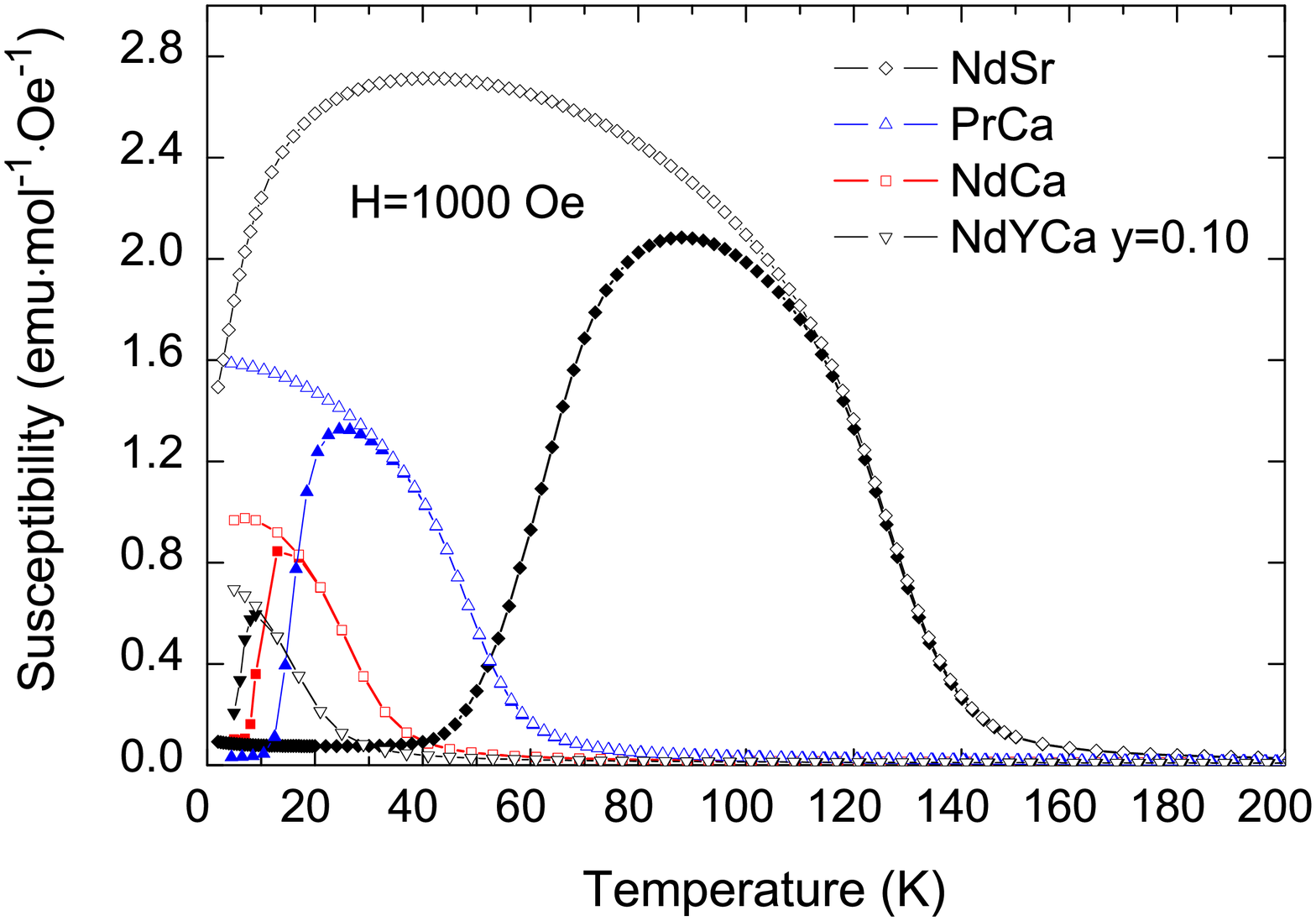}
\caption{(Color online) The zero-field-cooled (full symbols) and field-cooled (open symbols) curves of DC susceptibility in
\ndsrtri, \prcatri\ and \ndcatri, measured at field of 1000~Oe. The effect of an additional
disorder is demonstrated by the data of (Nd$_{1-y}$Y$_{y}$)$_{0.7}$Ca$_{0.3}$CoO$_3$ ($y=0.10$).}
 \label{fig3}
\end{figure}

The basic magnetic characterization of \ndsrtri, \prcatri\ and \ndcatri\ is represented by the ZFC
and FC susceptibilities $\chi_{ZFC}$, $\chi_{FC}$ measured in a field of 1000~Oe
(Fig.~\ref{fig3}). Starting from room temperature both susceptibilities increase with decreasing
temperature and the onset of FM phase (Curie temperature) can be specified from an inflection
point of the $\chi_{FC}$(T) dependence. This yields $T_C=130$, 55 and 25~K for \ndsrtri, \prcatri\
and \ndcatri, respectively. With further decreasing of temperature, the $\chi_{ZFC}$ curves
exhibit a maximum at a temperature decreasing with the increasing applied field. The $\chi_{FC}$
data for \prcatri\ increase steadily toward zero temperature, while those for \ndsrtri\ and
\ndcatri\ show a sudden decrease at the lowest temperatures. Such drop is caused by Nd$^{3+}$
moments that are induced by FM order in cobalt subsystem and orient antiparallel to
Co$^{3+}$/Co$^{4+}$ spins - see \textit{e.g.} the magnetic and neutron diffraction study of
\ndsrco\ sample with $x=0.33$ by Krimmel~\etal\ \cite{RefKrimmel2001PRB64_224404}. This effect is
not seen in \prcatri\ sample since the singlet ground state of Pr$^{3+}$ lacks intrinsic moment and shows only weak magnetic polarization that arises due to mixing with excited states.

\begin{figure}
\includegraphics[width=0.90\columnwidth,viewport=0 0 756 576,clip]{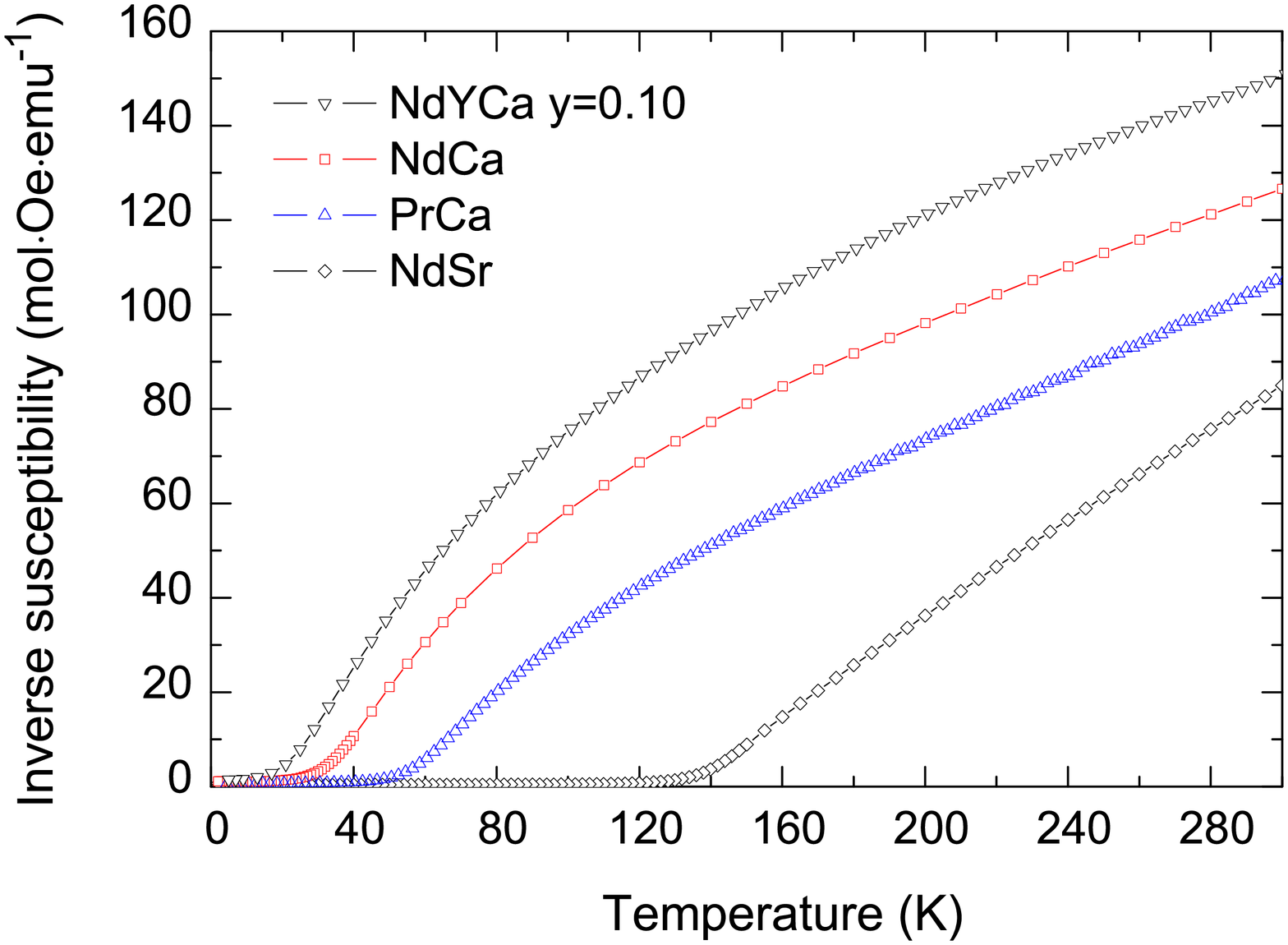}
\caption{(Color online) The inverse susceptibility for \ndsrtri, \prcatri, \ndcatri\ and
(Nd$_{1-y}$Y$_{y}$)$_{0.7}$Ca$_{0.3}$CoO$_3$ ($y=0.10$), based on FC data.}
 \label{fig4}
\end{figure}

The inverse susceptibility data are presented in a broader range of temperature in
Fig.~\ref{fig4}. The curvature above $T_C$, which is marked for \ndcatri\ and \prcatri\ but hardly
visible for \ndsrtri, represents a characteristic behaviour for samples with ferrimagnetic ground state.
This suggests that not only the Co-Nd but also Co-Pr exchange are of an antiferromagnetic (AFM) kind.

Taking into account the complexity of exchange interactions and crystal field effects, no definite
conclusions on the cobalt spin states and their temperature dependence can be drawn from the
low-temperature susceptibility data. A simpler situation occurs close to room temperature, where
the inverse susceptibility approaches a linear Curie-Weiss behaviour. The slope for \ndsrtri\
gives, after subtraction of free-ion value for Nd$^{3+}$ contribution, an effective moment of
$\mu_{eff}^2 \sim 10 \mu_B^2$ \textit{per} Co. This is exactly the theoretical value for IS/IS
Co$^{3+}$/Co$^{4+}$ mixture, which corroborates the results for \lasrco\ in the $300-600$~K range
reported by Wu and Leighton \cite{RefWu2003PRB67_174408}. On the other hand, the asymptotic
behaviour of high-temperature inverse susceptibility data for \prcatri\ and \ndcatri\ gives
notably larger effective moments $\mu_{eff}^2 \sim 15$ and 19 $\mu_B^2$ \textit{per} Co,
respectively, which may suggest that Co$^{3+}$ occur partially in HS state as it is for \laco\ at
the room temperature. This conjecture seems to be supported by observed Weiss temperature $\theta$
that changes from positive (FM, case of \ndsrtri) to negative (AFM, case of \prcatri\ and
\ndcatri) values. With larger disorder, the AFM interactions strengthen as clearly seen for the
\ndcatri\ sample with yttrium doping, added to Figs.~\ref{fig3} and \ref{fig4}. Although the AFM
interactions between cobalt and rare earths may play some role, we relate such drastic change of
magnetic interactions at high temperature mainly to the presence of excited HS Co$^{3+}$ states in
the matrix of dominant LS or IS Co$^{3+}$ character.

\begin{figure}
\includegraphics[width=0.90\columnwidth,viewport=0 0 567 804,clip]{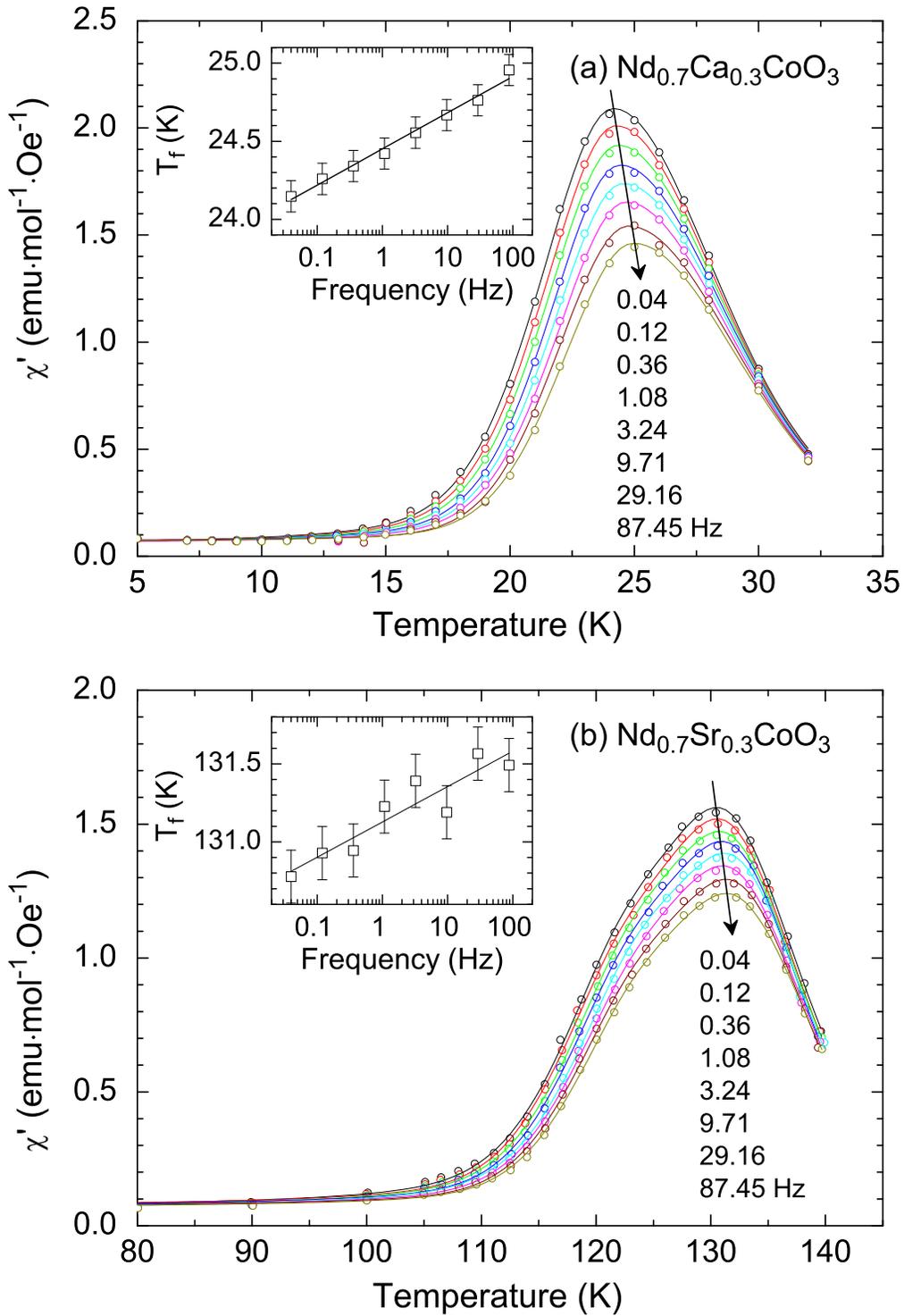}
\caption{(Color online) The real part of AC susceptibility for \ndcatri\ and \ndsrtri. The experimental data are marked by the symbols; the full lines represent the least-squares fit of the peak form. The inset shows the
frequency dependence of freezing temperature $T_f$, corresponding to the susceptibility maximum.}
 \label{fig5}
\end{figure}

The behaviour of \ndcatri\ and \ndsrtri\ in the vicinity of FM transition has been further probed
by AC susceptibility, for which real and imaginary parts culminate both near $T_C$. For \ndcatri\
it is clearly seen that the characteristic temperature $T_f$, at which the real part $\chi '$
passes through a maximum, exhibits an upward shift with increasing frequency $\nu$ of the applied
AC field (Fig.~\ref{fig5}). In analogy to freezing processes in the spin-glass systems, this shift
can be quantified by a semiempirical dimensionless parameter $K = \Delta T_f/[T_f \Delta
(log\nu)]$. The value of this parameter was evaluated from the linear approximation of the
dependence $T_f(log\nu)$ (see the inset of Fig.~\ref{fig5}) using the least-squares method, which
yields $K=0.0095\pm0.0005$. For \ndsrtri\ where frequency shift is less obvious, the analysis
gives $K=0.0017\pm0.0003$. The existence of a finite frequency shift means that some glassiness or
"glassy ferromagnetism" is involved in a broad range below $T_C$, and this refers not only to
weakly magnetic \ndcatri\ but also to \ndsrtri\ with stronger FM ground state.

\begin{figure}
\includegraphics[width=0.90\columnwidth,viewport=0 0 567 804,clip]{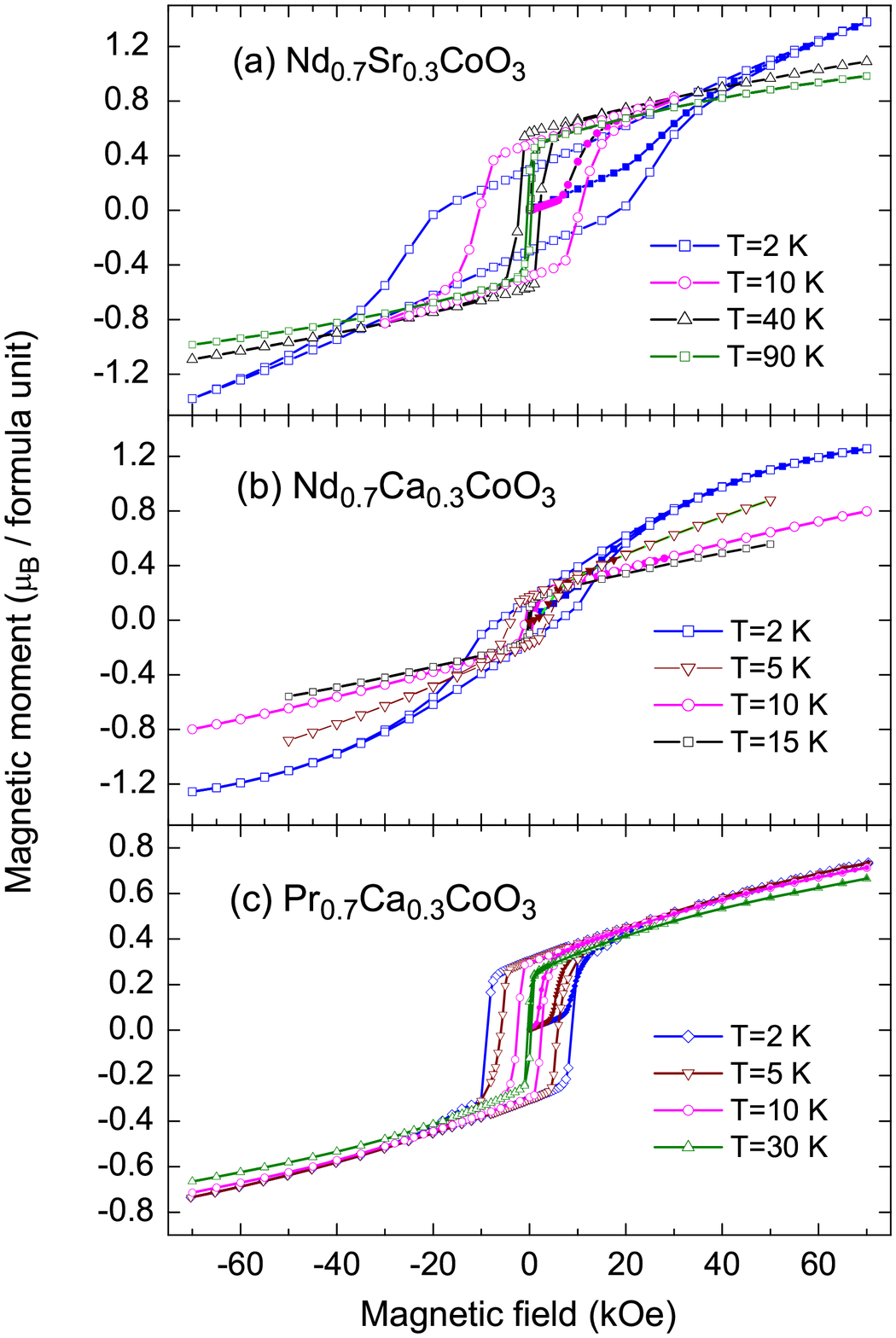}
\caption{(Color online) The virgin magnetization curves and ZFC hysteresis loops for \ndsrtri\
(a), \ndcatri\ (b) and \prcatri\ (c), taken at selected temperatures between $T_C$ and  2~K.}
 \label{fig6}
\end{figure}

The virgin magnetization curves and ZFC hysteresis loops in fields up to 70~kOe are presented in
Fig.~\ref{fig6}. The results at the lowest temperatures show a superposition of nearly rectangular
hysteresis loop with a linear component that is mostly due to rare-earth contribution. As expected
for the magnetic ground state of Kramers ions Nd$^{3+}$, this additional term (paraprocess) is
especially large for \ndsrtri\ and \ndcatri\ while it is much smaller and apparently
temperature-independent for \prcatri. The spontaneous magnetization of Co subsystem, estimated by
linear extrapolation to zero field, makes 0.68, 0.34 and 0.23 $\mu_B$/Co for \ndsrtri, \prcatri\
and \ndcatri, respectively. (In order to eliminate the contribution of Nd$^{3+}$ moments in
antiparallel alignment, the extrapolation was done using hysteresis curves at higher temperatures,
40~K for for \ndsrtri\ and 10~K for \ndcatri, and a small temperature correction of about 5\% was
applied, based rather arbitrarily on the $S=2$ Brillouin function.) Concerning the rare-earth
contribution we note that the magnetization curve for \ndcatri\ at 2~K tends in high fields to a
saturation of about 1.3~$\mu_B$ \textit{per} f.u., which should be interpreted as a sum of
$\mu$(Co)+0.7$\mu$(Nd). Since our Brillouin analysis of magnetization curves in a broader
temperature range did not reveal any observable paraprocess of Co subsystem in the \ndcatri\
sample, we may conclude that the contribution of neodymium moments makes about $1.0~\mu_B$ or
equivalently $1.4~\mu_B$ \textit{per} Nd$^{3+}$ ion.

The same Nd$^{3+}$ moment can be anticipated also for the \ndsrtri\ sample. Here, much higher
fields are needed to approach the magnetic saturation, because of stronger AFM coupling between Co
and Nd subsystems. Based on a trial measurement at 2~K up to 140~kOe, we have estimated the
saturated moment of 2.3~$\mu_B$ \textit{per} f.u. This may signify that, unlike in \ndcatri, the
cobalt moment in \ndsrtri\ increases notably under application of external field, namely from
0.67~$\mu_B$/Co at zero field to about 1.3~$\mu_B$/Co at 140~kOe. The different behaviour and
inhomogeneous character of the \ndsrtri\ sample has been also manifested in a shift of the center
of FC hysteresis loop toward negative fields (the exchange bias makes actually 230~Oe at $T=2$~K -
not shown in Fig.~\ref{fig6}a).

\subsection{The low-temperature specific heat}

\begin{figure}
\includegraphics[width=0.90\columnwidth,viewport=0 0 756 576,clip]{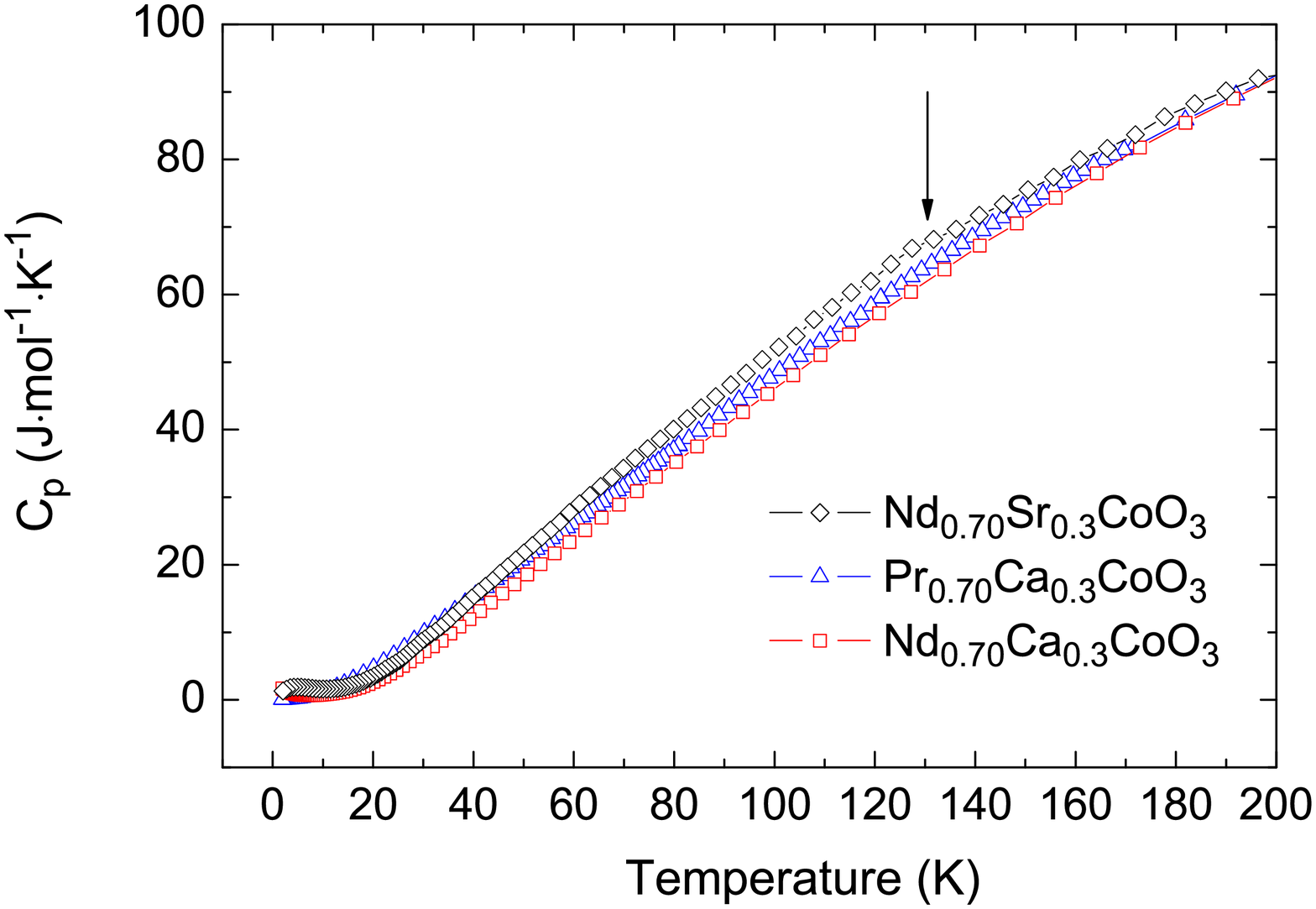}
\caption{(Color online) The heat capacity of \ndsrtri, \prcatri\ and \ndcatri\ at intermediate
temperatures. The arrow marks a weak anomaly at $T_C$ for \ndsrtri.}
 \label{fig7}
\end{figure}

\begin{figure}
\includegraphics[width=0.90\columnwidth,viewport=0 0 567 804,clip]{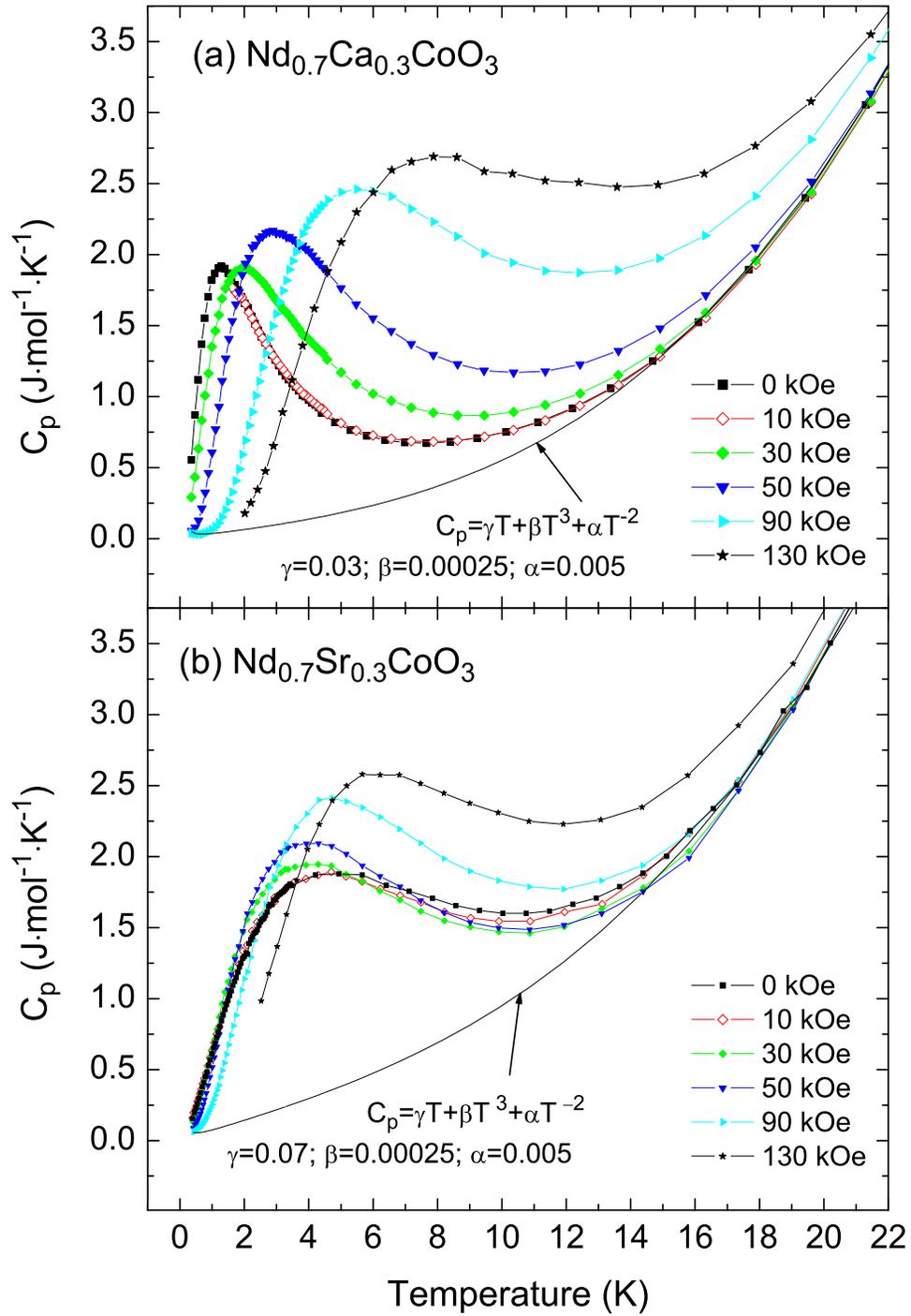}
\caption{(Color online) The low-temperature heat capacity of \ndcatri\ and \ndsrtri. The
background line corresponding to a contribution of the hyperfine, lattice and linear terms is
estimated with relative uncertainty of about 5\%, which reflects some ambiguity of the $\alpha$, $\beta$ and $\gamma$ parameters.}
 \label{fig8}
\end{figure}

The specific heat data for all studied samples are presented in a broad temperature range in
Fig.~\ref{fig7}. The main contribution comes from lattice dynamics, characterized by Debye
temperature $\theta_D \sim 350-400$~K for \ndsrtri, \prcatri\ and \ndcatri. In addition, there are
two contributions that may affect the specific heat values at intermediate temperatures. One is
the magnetic term due to cobalt ions, which is manifested by a very weak $\lambda$ reminding
anomaly at $T_C$ in \ndsrtri, but is supposedly spread over larger temperature range in \prcatri\
and \ndcatri. The second contribution is due to thermal excitation among the rare-earth 4$f$
electronic levels. In the \prcatri\ system with perovskite structure of the orthorhombic $Pbnm$
symmetry, the $^3$H$_4$ electronic multiplet of Pr$^{3+}$ is split by crystal field effects to
nine singlet levels that are displaced over a large energy range of 100 meV
\cite{RefPodlesnyak1994JPCM6_4099}. The thermal population of the first excitation level at about
5 meV is manifested in the specific heat as a broad Schottky-type contribution, onset of which is
readily seen as a hump in the \prcatri\ data at about 20 K. A more interesting situation is
encountered in \ndsrtri\ and \ndcatri\ where the $^4$I$_{9/2}$ multiplet of Nd$^{3+}$ is split to
five Kramers doublet levels \cite{RefPodlesnyak1993JPCM5_8973}. First, the energy gap between the
ground doublet and first excited doublet is larger, 12 meV, so that the relevant Schottky-type
contribution is shifted to higher temperatures and becomes more diffusive. Secondly, the presence
of internal field due to FM ordering in cobalt subsystem is responsible for lifting of Kramers
degeneracy of Nd$^{3+}$ electronic states by Zeeman effects. In particular, the ground level split
by internal field is a realization of a standard two-level system for pseudospins $J'=\pm$1/2
relevant to two eigenstates of the ground doublet. Their thermal redistribution is reflected by
appearance of characteristic Schottky peak in the low-temperature specific heat, see \textit{e.g.}
\cite{RefGordon1999PRB59_127,RefHejtmanek2010PRB82_165107}. For present samples, these Schottky
peaks and their shift with applied magnetic field are seen in more detail in Fig.~\ref{fig8}. In
the subKelvin range there is another field-dependent Schottky anomaly of the nuclear ($^{141}$Pr,
$^{143/145}$Nd, $^{59}$Co) origin. This $T^{-2}$ term that is especially strong for \prcatri\ (not
shown) because of large Van Vleck susceptibility of praseodymium in ground singlet state and
strong hyperfine coupling constant.

\begin{figure}
\includegraphics[width=0.90\columnwidth,viewport=0 0 567 804,clip]{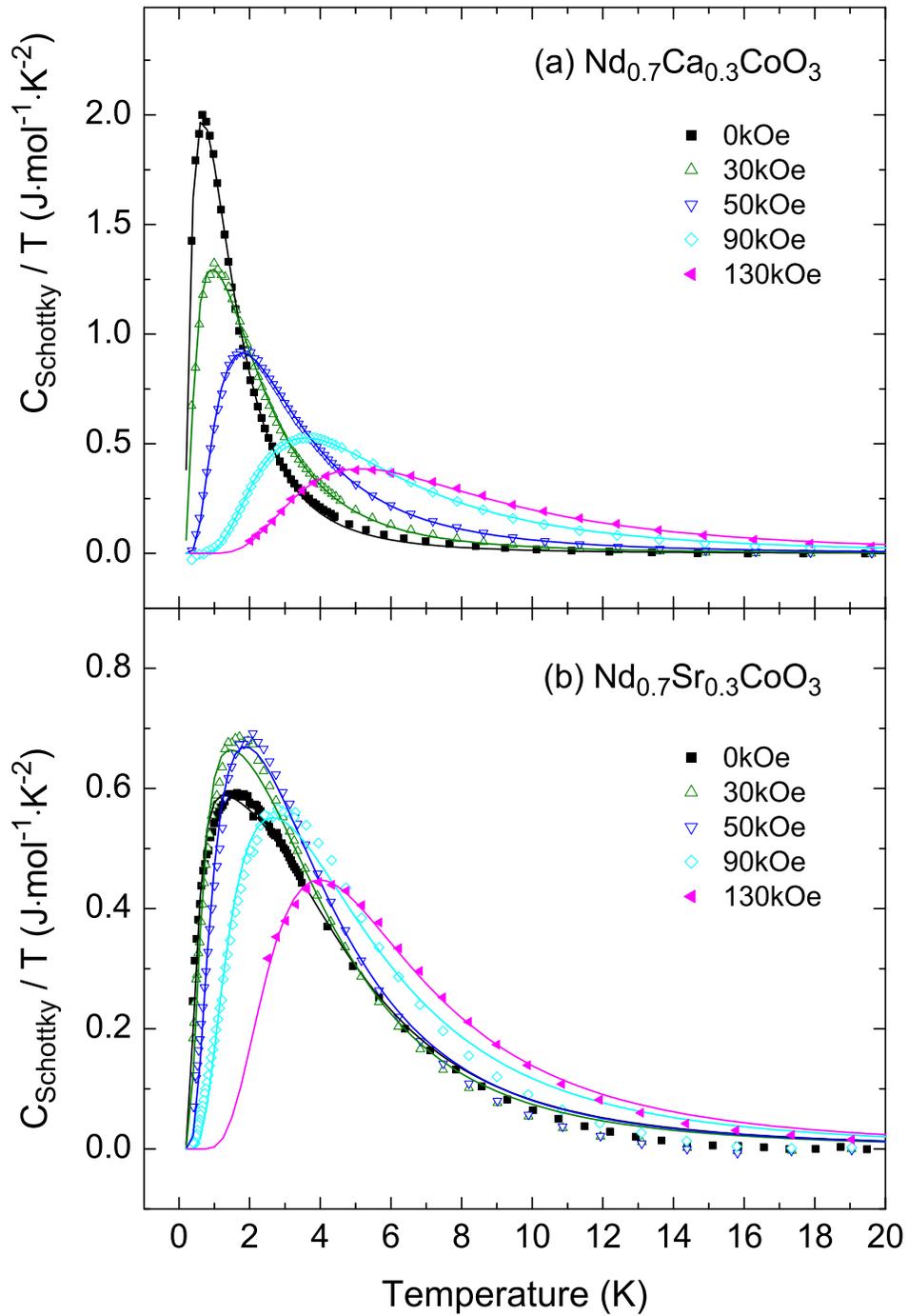}
\caption{(Color online) The heat capacity divided by temperature for \ndcatri\ and \ndsrtri, after
subtraction of lattice and nuclear terms. The full lines present the theoretical fit based on the
broadening due to anisotropic Zeeman splitting.}
 \label{fig9}
\end{figure}

\begin{figure}
\includegraphics[width=0.90\columnwidth,viewport=0 0 567 804,clip]{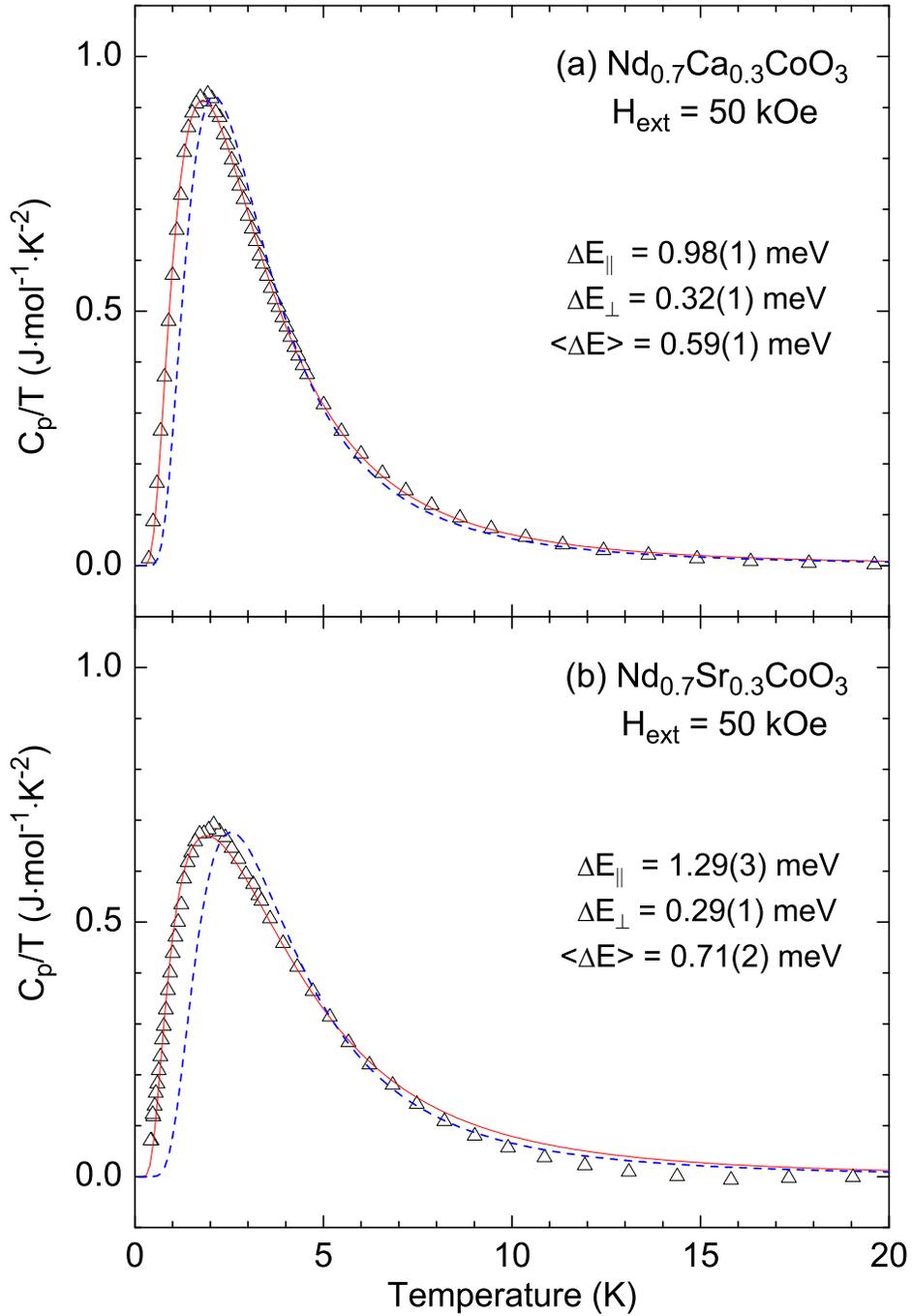}
\caption{(Color online) The $c_{Schottky}$ contribution to the low-temperature heat capacity of
\ndcatri\ and \ndsrtri\ at $H_{ext}=50$~kOe. The experimental peaks and their fitted profiles in model of anisotropic $\Delta E$ are compared with
the theoretical form for ideal Schottky peak (the dashed line), which would correspond to isotropic model with the same entropy change and splitting $<$$\Delta E$$>$.}
 \label{fig10}
\end{figure}

\begin{figure}
\includegraphics[width=0.90\columnwidth,viewport=0 0 756 576,clip]{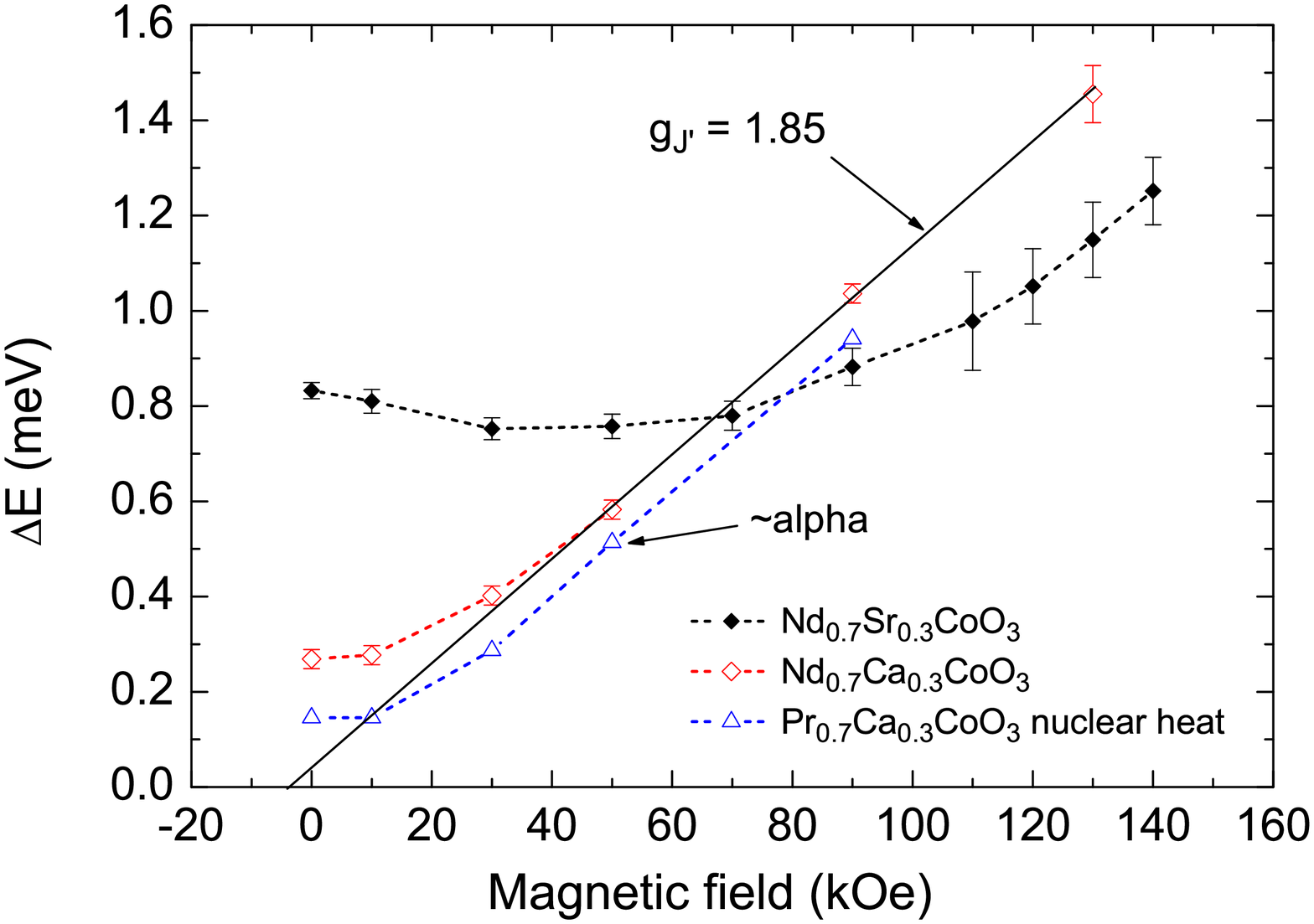}
\caption{(Color online) Average Zeeman splitting of the Nd$^{3+}$  ground doublet for \ndcatri\
and \ndsrtri. The linear behaviour for \ndcatri\ above $H_{ext}=50$~kOe gives
$<$$g_{J'}$$>$~=~1.85. Similar dependence is observed also for the nuclear term $\alpha.T^{-2}$ in
\prcatri. The parameter $\alpha$, which is proportional to hyperfine field induced by Van Vleck
susceptibility of the Pr$^{3+}$ electronic singlet, achieves the values
0.0034~J$\cdot$K$\cdot$mol$^{-1}$ and 0.0220~J$\cdot$K$\cdot$mol$^{-1}$ for $H_{ext}=0$ and
90~kOe, respectively.}
 \label{fig11}
\end{figure}

The profile of the Nd$^{3+}$ related Schottky peaks and location on the temperature scale
$\sim1-10$~K are controlled by Zeeman splitting of the ground doublet - $\Delta E = g_{J'}\mu_B
H_{eff}$, where $H_{eff}$ is the vector sum of internal and external fields experienced by
pseudospins $J'=\pm$1/2 in the solid state material. The relevant data, \textit{i.e.} after
subtraction of the hyperfine, lattice and linear terms, are displayed as $c_{Schottky}/T$ vs. $T$
in Fig.~\ref{fig9}. Although the  observed curves are broadened with respect to the ideal Schottky
peak as exemplified in Fig.~\ref{fig10}, no dual distribution of effective fields is found.
Macroscopic phase segregation is thus ruled out. Moreover, the observed broadening does not
necessarily mean inhomogeneous distribution of effective fields as it can be completely ascribed
to anisotropy of $g_{J'}$-factor and averaging in polycrystalline samples.  The analysis has been
actually made supposing an axial symmetry of $g$-tensor for the Nd$^{3+}$ ground doublet, so that
it is described by two components $g_\parallel$ and $g_\perp$ only. This model leads to a modified
Schottky form, where the energy splitting $\Delta E$ for a particular site is given by the angle
$\theta$ corresponding to the deviation of local $g_{J'}$-factor axis from the direction of the
magnetic field. The partial contribution to the overall Schottky-like anomaly is calculated as
[($\Delta E_{\parallel} cos\theta)^2$ + ($\Delta E_{\perp} sin\theta)^2]^{1/2}$ and the
contribution to specific heat is weighted by sin$\theta$, which corresponds to the random
orientation of crystallites in the sample. The fit, represented by solid lines in Figs.~\ref{fig9}
and \ref{fig10}, gives for \ndcatri\ the ratio $\Delta E_{\parallel}$/$\Delta
E_{\perp}$=$g_{\parallel}$/$g_{\perp} \sim 3.0$, irrespective the strength of applied field. The
\ndsrtri\ sample shows at $H_{ext}=0$ and 10~kOe an excessive broadening and lower height of
Schottky peaks. At higher fields, the curves acquire a similar form and the fit gives a ratio
$\Delta E_{\parallel}$/$\Delta E_{\perp}$=$g_{\parallel}$/$g_{\perp} \sim 4.5$. As to the absolute
values of Zeeman splitting and $g$-factors, the relevant data are plotted in Fig.~\ref{fig11}. It
is seen that average values of $\Delta E$ increase with external field in a gradual rate, and only
at higher fields a linear dependence is approached. Since the cobalt moments are practically
saturated in the Schottky peak range, the final slope seems to be a sole effect of the applied
field and can be thus used to determine the $g_{J'}$-factor. Average value actually obtained for
\ndcatri\ is $<$$g_{J'}$$>$=1.85. A similar field dependence is observed also in the plot of
nuclear contribution $\alpha T^{-2}$ for \prcatri, which probes Van Vleck polarization of
Pr$^{3+}$ ground singlet.

Turning back to the zero-field values of $\Delta E = g_{J'}\mu_B H_{eff}$ for \ndcatri\ and
\ndsrtri\ (see Fig.~\ref{fig11}), the spontaneous internal field acting on pseudospins can be
determined, using the value $<$$g_{J'}$$>$=1.85, to $\sim 25$ kOe and 80~kOe, respectively. The
last quantity reported here is the entropy change over the Schottky peak, calculated by
integration of $c_{Schottky}/T$. A value of $3.95\pm0.08$~\jkmol\ is obtained for \ndcatri, which
is 98\% of the theoretical value $0.7N k_B ln2=4.04$~\jkmol. For \ndsrtri, the value obtained from
zero-field data makes $3.45\pm0.07$~\jkmol\, and increases with increasing applied field to
$3.65\pm0.07$~\jkmol. This corresponds to 85\% and 90\% of the theoretical value, respectively.
The increase with applied field indicates an increasing number of the Nd$^{3+}$ pseudospins
contributing to Schottky peak. The reduced entropy value at zero field, the large width of
Schottky peak showing a gradual narrowing with the applied field, and the magnetic characteristics
mentioned above (paraprocess of Co subsystem, finite exchange bias) are signatures of an
inhomogeneous FM state in \ndsrtri.

\section{Discussion}

Our structural and electric transport data indicate, that the  \ndsrtri, \prcatri\ and \ndcatri\
samples are single-phase highly homogeneous polycrystalline systems of the perovskite $Pbnm$
structure with oxygen stoichiometry close to the ideal one. As the magnetic characteristics are
concerned, a comparison with available literature data on the same cobaltites is worthwhile. Both
the saturated magnetic moment 0.67~$\mu_B$/Co and $T_C=130$~K found for \ndsrtri\ sample are
definitely lower compared to 1.55~$\mu_B$/Co and 200~K, the data which were reported for
chemically most similar \ndsrco\ sample, nonetheless with declared slightly higher Sr content
$x=0.33$ \cite{RefKrimmel2001PRB64_224404}. Noting that this latter moment value is close to
1.71~$\mu_B$ achieved for \lasrtri, we presume  that the \ndsrco\ samples with  $x \sim 0.3$ are
close to the compositional transition from LS/LS groundstate to IS/LS one. This can explain the
large difference in magnetic response between these two samples, despite of their very similar
chemical composition. Considering two above mentioned possible ground states, the low critical
temperature $T_C=25$~K found for our \ndcatri\ sample can be understood as a fingerprint of a
single phase sample with LS/LS FM ground state, while a spin-glass freezing anomaly at much higher
temperature of about 55~K reported for the single crystal samples of same composition by
Kundu~\etal\ can reflect the admixture of the IS/LS ground state with significantly higher $T_C$
\cite{RefKundu2006JSSC179_923}. Simultaneously we rule out a drop of critical temperature due to
oxygen deficiency of our sample (see \textit{e.g.} Ref.~\cite{RefKundu2004JMMM281_261}), since in
spite of lower $T_C$, we observe a "stronger" FM state with spontaneous moment 0.23~$\mu_B$/Co and
significant coercivity of $\sim10$~kOe at 2~K.

More extensive data are available for the \prcatri\ system. Our sample, which oxygen stoichiometry
has been proved directly by the neutron diffraction, exhibits FM transition at $T_C=55$~K and the
spontaneous moment achieves 0.34~$\mu_B$/Co. Identical values and similar hysteresis loops were
reported for \prcatri\ also by Tsubouchi~\etal\ \cite{RefTsubouchi2004PRB69_144406}, while other
literature data cluster around higher $T_C=70$~K and the magnitude of bulk FM moment is generally
lower. The moment of 0.15~$\mu_B$/Co for \prcatri\ is reported in the above mentioned work of
Kundu~\etal\ The values found on \prcatri\ by El-Khatib~\etal\ are 0.20~$\mu_B$ as deduced from
magnetization measurements and the long-range ordered moment 0.30~$\mu_B$ \textit{per} f.u.
determined by neutron diffraction (this latter value is not explicitly mentioned in the paper, but
can be deduced from the graph of magnetic intensities). Cobalt moments of analogous value
$\sim$0.20~$\mu_B$ are obtained also by Kalinov~\etal\ \cite{RefKalinov2010PRB81_134427} on
\prcatri-related systems with small A-site substitution by Eu$^{3+}$ ions.
Importantly, the magnetization curves have been measured up to high fields of 140~kOe, and the
data evidence, after subtraction of Pr contribution, a saturation of cobalt magnetization of
$\sim0.35\mu_B$ \textit{per} f.u. This value is close to a sole contribution of Co$^{4+}$ ions and
evidences thus a dominance of LS/LS phase in systems \prcatri\ and consequently also \ndcatri.

The most significant results of present study refer to the heat capacity experiments, namely the
study and interpretation of the low-temperature Schottky peak associated with Nd$^{3+}$ ground
state doublet. Using this approach we have demonstrated that presence of rare-earth ions with
Kramers degeneracy can be used as a local magnetic probe in mixed-valence cobaltites. Beyond the
quantitative information on entropy and level splitting, it is also possible to analyze the
intrinsic or inhomogeneous broadening of the corresponding low temperature Schottky peaks. For
this purpose we have performed the theoretical calculations based on the parameters of crystal
field, which have been recently deduced from extensive study of terbium aluminate TbAlO$_3$
possessing the same $Pbnm$ perovskite structure \cite{gruber}.  This procedure, described in
detail in Appendix, provides not only the necessary characteristic energies of energy level
splitting but also the actual form of respective doublet or singlet states. In addition to $4f^3$
electronic configuration of Nd$^{3+}$ and $4f^2$ for Pr$^{3+}$, the calculations are done also for
the $4f^1$ electronic configuration, which is relevant for the cases of Ce$^{3+}$ or Pr$^{4+}$.
Turning back to the $^4$I$_{9/2}$ multiplet of Nd$^{3+}$, the ground doublet is characterized by
highly anisotropic $g$-tensor with principal components $g_x=4.472$, $g_y=1.185$ and $g_z=0.928$
or, in pseudoaxial approximation $g_\parallel$/$g_\perp \sim 4.2$ (see Table III in Appendix). For
enabling a comparison with our experiments, a numerical integration over random orientation of the
crystallites has been made, yielding the average value $<$$g_{J'}$$>=2.51$. This corresponds to
the pseudospin moment $1.255~\mu_B$, which is in reasonable agreement with the above mentioned
saturated magnetization in \ndcatri, giving an estimate of $1.4~\mu_B$ \textit{per} Nd$^{3+}$ ion.
On the other hand, the value $<$$g_{J'}$$>=1.85$ deduced from the high-field shift of Schottky
peaks is much lower. An opposite discrepancy has been observed for another Kramers ion Pr$^{4+}$
in the low-temperature phase of \prycatri\ with $y=0.15$, where shift of Schottky peaks gives
unexpectably large $<$$g_{J'}$$>=3.30$ \cite{RefHejtmanek2010PRB82_165107}, while calculations in
Table II in Appendix give $<$$g_{J'}$$>=2.07$ and a still lower value would be obtained if
correction for large Pr$^{4+}$ covalency were made. These findings may suggest that apart of the
bare interaction of external field with $4f$ moments, there is an indirect interaction through
spin polarizable electronic cloud. The same mechanism likely mediates also the exchange
interaction between FM ordered cobalt subsystem and Kramers pseudospins for rare earths, which is
of AFM type for Nd$^{3+}$ and FM type for Pr$^{4+}$ \cite{RefNoteA}.

\section{Summary}

The studied mixed-valence cobaltites are systems with complex behaviour. These compounds possess
an intrinsic inhomogeneity that relates to chemical and size disorder at the perovskite A-sites,
possibility of various spin states at cobalt sites and macroscopic distortion associated with
cooperative tilt of CoO$_6$ octahedra. With increasing octahedral tilt (decreasing Goldschmidt's
tolerance factor) the FM interactions are suppressed. In particular for \lasrtri, \ndsrtri,
\prcatri\ and \ndcatri\ with the same 30\% doping level, the magnetic Curie temperature gradually
decreases,  $T_C=230$~K, 130~K, 55~K and 25~K, and the spontaneous FM moment on cobalt sites
drops, 1.71~$\mu_B$, 0.68~$\mu_B$, 0.34~$\mu_B$ and 0.23~$\mu_B$. A difference is also in critical
behavior around the magnetic transitions. The \lasrtri\ system behaves as conventional
ferromagnet. The systems with reduced magnetization show frequency dependent AC susceptibility
peaks, pointing to a presence of larger FM clusters for \ndsrtri\ and smaller clusters for
\ndcatri\ or \prcatri\ in certain temperature range below $T_C$. (A more direct evidence of
unusual behaviour in latter systems comes from small angle neutron scattering on \prcatri, which
shows a presence of preformed FM entities of 10~\AA\ size, that only well below $T_C$ grow to
macroscopic size \cite{RefElKhatib2010PRB82_100411}.)

The magnetization data on \ndsrtri, \prcatri\ and \ndcatri\ at low temperatures exhibit standard
FM hysteresis loops combined with paraprocess caused by rare-earth paramagnetism. The long-range
character of magnetic ordering is reported also in few neutron diffraction experiments, which
however cannot decide whether severely suppressed moment in \prcatri\ and \ndcatri\ is a
manifestation of uniform phase of weakly magnetic character or refers to two-phase coexistence of
macroscopic FM and non-magnetic regions \cite{RefNoteB}. Using the low-temperature heat capacity
experiments as an efficient tool for analyzing of internal magnetic fields, we decide in favour of
essentially homogeneous phase in the calcium based compounds down to nanoscopic or even atomic
scale. First we note that the magnitude of cobalt moments is remarkably close to 0.3~$\mu_B$,
which is exactly the theoretical value for a mixture of non-magnetic $S=0$ ions LS Co$^{3+}$ and
$S=1/2$ ions LS Co$^{4+}$. No metamagnetic increase of cobalt magnetization has been detected on
application of fields up to 140~kOe. This strongly supports our conclusion that the ground state
is the LS/LS phase with only minor promotion of $t_{2g}$ electrons to itinerant $e_{g}$ levels.
This promotion is necessarily associated with a rise of cobalt moments, but the increased
magnetization might be compensated by presence of oppositely oriented impurity moments due to
isolated $S=2$ ions HS Co$^{3+}$. Secondly, the form and intensity of Schottky peaks originating
in Zeeman splitting of the ground state Kramers doublet of Nd$^{3+}$ show unambiguously that all
rare-earth sites experience the same effective field. The uniform distribution of hole carriers
(formally LS Co$^{4+}$) in the main LS/LS phase is thus firmly established for \ndcatri\ and can
be anticipated for \prcatri, as well, in order to account for the similarly reduced magnetization
of Co subsystem $\sim0.3 \mu_B$. For the remaining compound \ndsrtri, the observed magnetization
is intermediate between those for LS/LS in \ndcatri\ and IS/LS in \lasrtri. Based on a reduced
intensity of the Nd$^{3+}$-related Schottky peak, we infer that some rare-earth sites are located
in magnetically disordered regions or, to account for exchange bias in \ndsrtri, in AFM ordered
regions. Such finding is a direct evidence of inhomogeneous state of this sample.

As a final remark, let us note that present results are closely related to a general problem of
the $3d-4f$ exchange in perovskite oxides, mediated presumably by spin polarization of extended
orbitals of the rare-earth $5d$ parentage - see the Campbell's indirect exchange mechanism treated
in \cite{RefRichter1998JPDAP31_1017} and references therein. It is not yet clear how the mixed
valence character of cobaltites, \textit{i.e.} the presence of itinerant  carriers, affects the
strength and eventually also the sign of exchange interaction. To elucidate all these issues, the
magnetic and heat capacity studies on twin-free single crystals are desirable. The single crystal
experiments may also give a better test of the homogeneity of FM phases, probed by analysis of the
actual form of Schottky peaks, since excessive broadening due to anisotropic $g$-factors will be
eliminated.

\appendix

\section{Electron states and magnetism of lanthanide ions in orthorhombic perovskites}

To describe the $4f$ states of lanthanide ions a Hamiltonian that consists of the free-ion
(atomic) and crystal field terms is routinely used:
\begin{equation}
\label{eq:h}
 \hat{H}=\hat{H}_a + \hat{H}_{CF}.
\end{equation}

The free ion Hamiltonian is spherically symmetrical and in a standard notation (see, for example,
Ref. \cite{carnall}) it can be written as
\begin{eqnarray}
\label{eq:hatom}
 \hat{H}_a = E_{avg} + \sum_{k=2,4,6}F^k \hat{f}_{k} + \zeta_{4f} \sum_{i=1}^{N} \hat{s}_{i} \hat{l}_{i} + \;\;\\
 \nonumber \alpha \hat{L}^2 + \beta\hat{G}(G_2) + \gamma \hat{G}(R_2) +  \;\;\\
 \nonumber \sum_{j=0,2,4}M^j\hat{m}_j + \sum_{k=2,4,6} P^k \hat{p}_k + \sum_{r=2,3,4,6,7,8}T^r
\hat{t}(r),
\end{eqnarray}
where $E_{avg}$ is the energy in central field, terms proportional to $F^k$, $\alpha, \beta,
\gamma$ and $T^r$ describe the electron-electron interaction, terms with $\zeta_{4f}$, $M^j, P^k$
parameters represent the spin-orbit, spin-other-orbit and electrostatically correlated spin-orbit
interactions. $N$ is number of the $4f$ electrons.

Within a single electron, crystal field theory the crystal field Hamiltonian may be written as
\cite{wybourne}
\begin{equation}
 \hat{H}_{CF} = \sum_{k,q,i} B_{q}^{(k)} \hat{C}_{q}^{(k)}(i),
\label{eq:hcf}
\end{equation}
where $ \hat{C}_{q}^{(k)}(i)$ is a spherical tensor operator of rank $k$ acting on $i$th electron
and the summation involving $i$ is over the $f$ electrons of lanthanide ion. $B_{q}^{(k)}$ are
crystal field parameters, the values of $q$ and $k$ for which they are nonzero depend on the site
symmetry and also on the choice of the local coordinate system. The local symmetry of the
lanthanide site in orthorhombic perovskites is $C_s$ and choosing the crystal field  coordinate
axes  along the orthorhombic axes $a,\,b,\,c$ results in three nonzero, real $B_{0}^{(k)}$
parameters ($k$ = 2, 4, 6) and six nonzero complex $B_q^{(k)}$ ($k = 2,4,6; q=2,4,6;\; q\leqq k$)
parameters. Rotation of the crystal field coordinate system around the $c$ axis allows elimination
of the imaginary part of the $B_2^{(2)}$ parameter (for detailed discussion of the crystal field
in orthorhombic perovskites see Ref. \cite{gruber}). The crystal field axes $x,\, y,\, z$, ( $z
\parallel c$) obtained in this way are also principal axes of the lanthanide susceptibility and
$g$ tensor and this system is used in what follows. Low symmetry of the crystal field leads to a
complete lift of the orbital degeneracy of $4f$ levels, so that the states are either singlets
($N$ even) or Kramers doublets ($N$ odd).

For the free lanthanide ions the total angular momenta $L,\;S,\;$ and $J$ are good quantum
numbers. The ground states are in accord with the Hund's rules and for Ce$^{3+}$ ($f^1$),
Pr$^{3+}$ ($f^2$) and Nd$^{3+}$ ($f^3$) they are $^2F_{5/2}$ ($S$=1/2, $L$=3, $J$=5/2), $^3H_{4}$
($S$=1, $L$=5, $J$=4) and $^4I_{9/2}$ ($S$=3/2, $L$=6, $J$=9/2), respectively. For a given
lanthanide ion the values of parameters of the atomic Hamiltonian (\ref{eq:hatom}) depend to some
extent on the host compound. This dependence has little significance for the results given below,
however. Decisive for the low temperature behavior of the lanthanide ions is thus the crystal
field.

In order to calculate the electron states of trivalent lanthanides in orthorhombic perovskites we
used the program 'lanthanide' \cite{edvardsson}, which makes possible to determine energy levels
and eigenfunctions of Hamiltonian (\ref{eq:h}) with external magnetic field added. For the atomic
parameters values given by Carnall~\etal\ \cite{carnall} were adopted. The crystal parameters
$B_q^{(k)}$ were taken to be the same as those determined recently by Gruber~\etal\ \cite{gruber}
for the non-Kramers Tb$^{3+}$ ion in terbium aluminate. Eigenenergies and magnetic moments we
calculated for the $4f^8$ electron configuration practically coincides with those obtained by
Gruber~\etal\ \cite{gruber}. In particular the energy difference between the lowest two  Tb$^{3+}$
singlets is only 0.026 meV, very large magnetic moment (8.8 $\mu_B$ at magnetic field 50~kOe) has
an Ising-like character with the Ising axis parallel to the $x$ axis of the crystal field. There
are two crystallographically equivalent sites of Tb$^{3+}$ ions in the unit cell related by
reflection in the $ac$ plane. The axis $x$ for these two sites makes an angle $\pm$ 36$^o$ with
the orthorhombic axis $a$.

\begin{figure}
\includegraphics[width=0.90\columnwidth,viewport=0 0 756 576,clip]{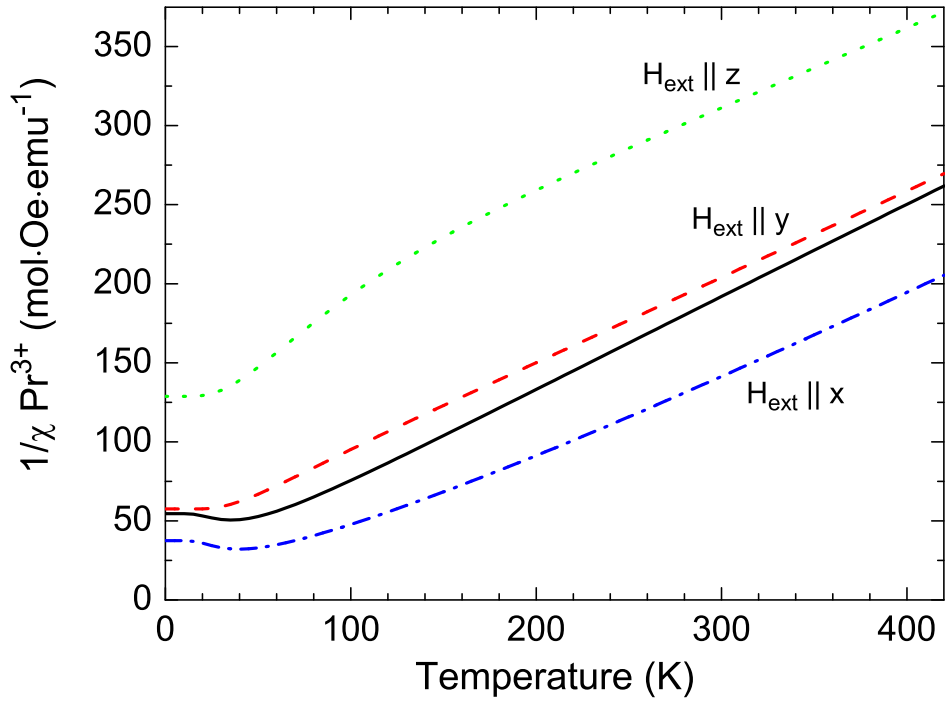}
\caption{(Color online) Theoretical dependence of the inverse susceptibility of Pr$^{3+}$ ion,
based on the singlet state energies in Table I and the calculated shifts in field of 10~kOe. The full line corresponds to the average over random orientations; the linear Curie behavior at higher temperatures gives an effective moment of $\mu_{eff}=3.68\mu_B$ in agreement with the experimental value $\sim 3.5\mu_B$.}
 \label{fig:chiprt}
\end{figure}

Compared to Tb$^{3+}$ the Pr$^{3+}$ ion is less anisotropic. The energies and magnetic moments of
the states originating from the lowest $^3H_{4}$ multiplet of Pr$^{3+}$ are given in Table I. The
dependence of magnetic moments on magnetic field is to a good approximation linear, only for the
first and second excited states and the field parallel to the $x$ axis there is a tendency to a
saturation in high magnetic fields. The induced magnetic moments are small with exception of the first and
second excited states that may be classified as Ising-like pseudodoublet. The susceptibility $\chi$ is anisotropic
and its temperature dependence is displayed in Fig.~\ref{fig:chiprt}.

Ce$^{3+}$ and Nd$^{3+}$ are Kramers ions with $4f^1$ and $4f^3$ electron configuration
respectively.  Magnetic moments of all Kramers doublets are almost field independent: they
increase by 1-5\% when field changes between 10 and 100~kOe. There is considerable anisotropy of
the moment, the $x$ axis being the easy axis of the ground doublet for both Ce$^{3+}$ and
Nd$^{3+}$. The $g$ factors of the Kramers doublets may be determined by multiplying the magnetic
moments by factor two, corresponding to a pseudospin 1/2. The calculated energies and the $g$
factors are summarized in Tables II and III.

\textbf{Acknowledgments}. This work was supported by Project No.~204/11/0713 of the Grant Agency
of the Czech Republic.


\begin{thebibliography}{}
\bibitem{RefNoteA}
The exchange interaction couples cobalt spins with the spin component of rare-earths moments, which is oppositely oriented to the total momentum in the $J=L-S$ multiplet. The antiparalel orientation is valid also for projection of spin momentum in Kramers pseudospin~$J'$.
\bibitem{RefNoteB}
In the theoretical case of FM-nonFM phase coexistence, the average value of ordered moment deduced from neutron diffraction is higher than average value deduced from magnetization data (this follows from the fact that magnetic diffraction intensities give a root-square-mean value of moments). This situation seems to occur for \prcatri\ sample probed by El-Khatib~\etal, but the value 0.3~$\mu_B$ (compared to 0.2~$\mu_B$ obtained by magnetization method) is close to the resolution limit of neutron diffraction and is thus subjected to large uncertainty.
\end{thebibliography}

\newpage

\begin{table}
\caption{States of the ground $^3H_{4}$ multiplet of Pr$^{3+}$ ion split by the crystal field.
Energy, relative to the ground state, at zero external magnetic field, and the magnetic moments
$m_x$, $m_y$, $m_z$, induced by field of 10~kOe.} \label{tab:pr}
\begin{tabular}{|c|c|c|c|c|}
\hline
state  & $E$ [meV] & $m_x$ [$\mu_B$]& $m_y$ [$\mu_B$]& $m_z$  [$\mu_B$]\\
\hline
   1 &     0.00 &     0.0469 &     0.0308 &     0.0139 \\
   2 &     8.29 &     0.2765 &     0.0536 &    -0.0021 \\
   3 &     9.94 &    -0.2484 &    -0.0182 &     0.0514 \\
   4 &    11.57 &    -0.0667 &    -0.0302 &    -0.0431 \\
   5 &    27.57 &     0.0620 &     0.0321 &     0.0144 \\
   6 &    29.98 &    -0.0624 &    -0.0407 &    -0.0038 \\
   7 &    52.92 &     0.0073 &     0.0041 &    -0.0216 \\
   8 &    63.61 &    -0.0052 &    -0.0250 &     0.0743 \\
   9 &    74.20 &    -0.0081 &    -0.0043 &    -0.0811 \\
\hline
\end{tabular}
\end{table}

\begin{table}
\caption{ Three Kramers doublets of the ground $^2F_{5/2}$ multiplet of Ce$^{3+}$ ion split by the
crystal field. Energy, relative to the ground state, at zero external magnetic field, and the $g$
factors.} \label{tab:ce}
\begin{tabular}{|c|c|c|c|c|}
\hline
state  & $E$ [meV] & $g_x$& $g_y$ & $g_z$\\
\hline
 1  &    0.00 &     3.757 &     0.935 &     0.606 \\
 2  &   33.46 &     1.298 &     2.298 &     1.458 \\
 3  &   56.36 &     0.945 &     1.212 &     3.451 \\
\hline
\end{tabular}
\end{table}

\begin{table}
\caption{Five Kramers doublets of the ground $^4I_{9/2}$ multiplet of Nd$^{3+}$ ion split by the
crystal field. Energy, relative to the ground state, at zero external magnetic field, and the $g$
factors.} \label{tab:nd}
\begin{tabular}{|c|c|c|c|c|}
\hline
state  & $E$ [meV] & $g_x$& $g_y$ & $g_z$\\
\hline
 1  &    0.00 &     4.472 &     1.185 &     0.928 \\
 2  &   11.83 &     0.942 &     3.925 &     1.077 \\
 3  &   22.87 &     2.034 &     1.257 &     3.540 \\
 4  &   53.52 &     2.978 &     3.063 &     1.052 \\
 5  &   65.33 &     1.788 &     1.435 &     4.204 \\
\hline
\end{tabular}
\end{table}

\end{document}